# Deep Phase Shifter for Quantitative Phase Imaging

QINNAN ZHANG,[1] SHENGYU LU,[1] JIAOSHENG LI,[3] WENJIE LI,[1] DONG LI,[1] XIAOXU LU,[3] LIYUN ZHONG[3] AND JINDONG TIAN[1,2, *]

[1]*College of Physics and Optoelectronic Engineering, Shenzhen University, Shenzhen 518060, China*
[2]*Guangdong Laboratory of Artificial Intelligence and Digital Economy (SZ), Shenzhen University*
[3]*Guangdong Provincial Key Laboratory of Nanophotonic Functional Materials and Devices, South China Normal University, Guangzhou 510006, China*
*\*jindt@szu.edu.cn*

**Abstract:** A single intensity-only holographic interferogram can records the full amplitude and phase information of optical field. However, current digital holography technologies cannot recover the lossless phase information from a single interferogram. In this paper, we provide an entirely new approach for the full-field quantitative phase imaging technology. We demonstrate that deep learning can be used to replace the entitative phase shifter, and quantitative phase imaging can obtain quantitative phase from a single interferogram in in-line holography. A deep-phase-shift network (DPS-net) is reported, which can be trained with simulation training data. The trained DPS-net can be used to generate multiple interferograms with arbitrary phase shift from a single interferogram as an artificial intelligence phase shifter. The ability and the accuracy of generating arbitrary phase shifts are verified, and the performance of the proposed method is also verified by the experimental interferogram. The results demonstrate that the proposed method can provide a full digital phase shifter with high-accuracy for the technology of dynamic quantitative phase measurement.

## 1. Introduction

Quantitative phase imaging (QPI) as a full-field, noncontagious and label-free quantitative imaging technology, is becoming a valuable method for investigating cells and tissues, industrial inspection, and study of hydrodynamics, because its nanoscale sensitivity to morphology and dynamics, and lower phototoxicity and no photobleaching imaging of completely transparent samples based on intrinsic contrast [1-6]. As we all know, digital holography techniques can recover the full-field phase information from intensity-only measurements. And various methods have been proposed to retrieval the phase information [7-16]. The current methods can be classified mainly in two types: one is off-axis digital holography, which can retrieval the phase information from a single interferogram [8-10]. It has been widely used in dynamic measurement. However, the introducing spatial carrier frequency will affect the measurement of high frequency information, and the space-bandwidth product of microscopy imaging system cannot be fully utilized, due to the use of off-axis incident field.

  The other one is phase shifting digital holography, which can precisely retrieval the phase information from more than one interferograms. It has been widely used for precision measurement. Various temporal or spatial phase shifting methods have been used to realize the recording of multiple phase shifting interferograms [11-15]. However, due to the inherent defects of temporal and spatial phase shifting, phase shifting technologies are difficult for dynamic measurement, and make the system more complicated and error-prone. In recent years, a digital four-step phase-shifting technique was proposed, which used $n$th-order Riesz transform components to combine three $\pi/2$ shifted interferograms [16]. However, the Riesz transform as the two-dimensional extension of the Hilbert transform [17,18], it has requirements for the frequency of sample information. And, it only generates three π/2 shifted interferograms. Strictly speaking, it does not replace the entitative phase shift devices.

Here, we propose a full digital phase shifter, name as deep phase shifter, which can be used to generate multiple interferograms with arbitrary phase shift from a single interferogram. Deep learning as a powerful method has attracted broad attention. Artificial neural networks can be trained to predict some measurable data. Therefore, deep learning is becoming a useful method for solving traditional problems in several optical fields, such as microscopy [19-22], image reconstruction [23-29], and super-resolution single-molecule microscopy [30]. In this paper, we report a deep-phase-shift network (DPS-net), which can be trained using simulation data set. After its training, the trained DPS-net can be used as a phase shifter. And conventional phase shift algorithms can be used to realize phase recovery. The following simulation and experimental results demonstrate the feasibility and precision of our method. It demonstrates that deep learning can be used to replace the entitative phase shifter, and improve the full-field quantitative phase imaging technology.

## 2. Methods

*Sample Preparation*: Hemispherical polyvinyl chloride pellets and Hela cells are used as the test samples. Hemispherical polyvinyl chloride pellets are dripped on a coverslip. And 100 uL HeLa cells with the density of $1\times10^5$ cells/mL were cultured in a coverslip, and dulbecco's modified eagle medium (DMEM) containing10% fetal bovine serum (Hyclone, Jackson Immuno Research, West Grove, PA, USA), and 1% of the antibiotics Penicillin G (100 units/mL)/Streptomycin (100 μg/mL) (Gibco, Jackson Immuno Research,West Grove, PA, USA) in the incubator with the temperature of 37 °C and $CO_2$ of 5 % concentration for 3 days.

*Holographic Imaging*: A Mach-Zehnder interference system is used in the experimental system, shown as in Fig. 1, a camera was used for recording the interferogram intensity distribution, an objective (20×, *NA*:0.4) was used for imaging the sample on the camera with 1024 × 768 pixels, and the pixel size was 10 μm. The wavelength of the laser was 632.8 nm.

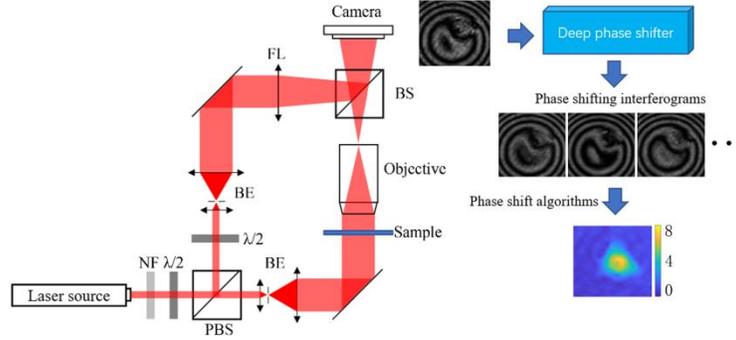

Fig. 1 The experimental microscopic interference system. NF: Neutral filter, λ/2: Half wave plate, BS: Beam splitter, BE: Beam expander, FL: Fourier lens.

*Deep Phase Shifter*: Deep phase shifter is a full digital phase shifter, which is set up based on neural networks. By using the single-channel initial interferogram as the input and the shared parameters in the network, we can output multi-channel phase shifting interferograms simultaneously. And the phase shifts in the interferograms for each channel can be controlled independently. In this paper, a DPS-net is trained to predict interferograms with arbitrary phase shifts from initial interferogram as the deep phase shifter. Its architecture is based on U-net [31]. The architecture is shown as in Fig. S1. (in Supplement 1). The details of network architecture are described as in Supplement 1.

*Generating training data*: Simulation training data are used for training, and input initial interferograms and the label interferograms are calculated using Eq. (1).

$$I_i(x, y) = A(x, y)^2 + B(x, y)^2 + 2A(x, y)B(x, y)\cos[\varphi(x, y) + \phi(x, y) + \Delta\phi_i] , \qquad (1)$$

where $\Delta\phi_i \, (i=0,1,2\cdots)$ is the phase shift, $\Delta\phi_0$ is zero, the corresponding interferogram is used as the initial interferogram. $\Delta\phi_i \, (i=1,2,3\cdots)$ are the phase shifts, we can define as any value we want, and the corresponding interferograms are used as the label. $A(x, y)$ and $B(x, y)$ represent the reference light and object light amplitude, and they are calculated as follow equation

$$A(x,y) = a_1 \times \exp\left[-(x-x_1)^2 / 2a_2^2 - (y-y_1)^2 / 2a_2^2\right] + \eta(x,y), \tag{2}$$

$$B(x,y) = b_1 \times \exp\left[-(x-x_1)^2 / 2b_2^2 - (y-y_1)^2 / 2b_2^2\right] + \eta(x,y), \tag{3}$$

where $a_1$, $b_1$, $a_2$, $b_2$, $x_1$ and $y_1$ are random number, respectively. The $\eta(x,y)$ is a zero-mean-value Guassian random noises, $\varphi(x,y)$ denotes the measured phase, $\phi(x,y)$ denotes the background phase, as $\phi(x,y)=2\pi n((x+a)^2+(y+b)^2)/1.28^2+\theta$, where $n$, $a$ and $b$ are random number; the $\theta$ is set to a value within the 0–2π range; the measured phase is obtained by the linear superposition of $k$ two-dimensional Gaussian functions, and can be represented as

$$\varphi = \sum_{k=1}^{K} \omega_k \exp\left[\left(\frac{x-\mu_x}{\sigma_x}\right)^2 - \frac{2\rho(x-\mu_x)(y-\mu_y)}{\sigma_x \sigma_y} + \left(\frac{y-\mu_y}{\sigma_y}\right)^2\right], \tag{4}$$

## 3. Results and discussion

### Training

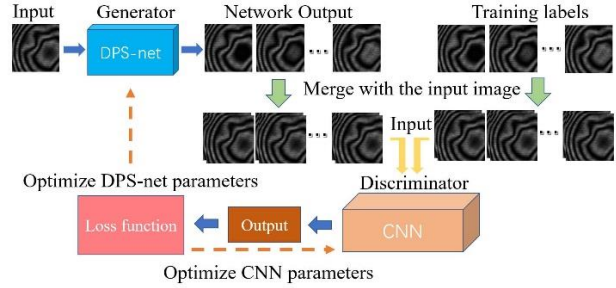

Fig. 2 Schematics of the DPS-net trained for phase shift

To train the DPS-net, the training strategy of generative adversarial network (GAN) is used [32]. DPS-net is used as the generator. A convolutional neural network (CNN) is used as the discriminator, the architecture is shown as in Fig. S1. (in Supplement 1). The schematics of the DPS-net trained for phase shift is shown as in Fig. 2. The initial interferogram is used as the network input. Firstly, by using the generator, multiple generative phase shifting interferograms can be obtained as the network output. Afterwards, two networks are trained according to the following conditions. 1) If the generative phase shifting interferograms and the corresponding label images are input the discriminator; we optimize the generator network. 2) If two group identical label images are input the discriminator; we optimize the discriminator network. Game Theory is used to train minimizes the loss functions. The corresponding loss functions can be represented as

$$Loss(G) = \sum_{x \times y} \frac{\log\left[1 + \exp(-D(G(x)))\right]^{-1}}{N_x \times N_y} + 10^2 \times R(G(x), X) + 10^4 \times \Delta\theta, \tag{5}$$

$$Loss(D) = \sum_{x \times y} \frac{\log\left[1+\exp(-D(X))\right]^{-1}}{N_x \times N_y} + \sum_{x \times y} \frac{1}{N_x \times N_y} \log\left[1 - \frac{1}{1+\exp(-D(G(x)))}\right], \quad (6)$$

where $Loss(G)$ is the loss function of the generator network, $Loss(D)$ is the loss function of the discriminator network, $D(\cdot)$ is the output of discriminator network, $G(x)$ is the generated interferograms, and $X$ is the corresponding label interferograms. $R(G(x), X)$ is the root-mean-square error (RMSE) between the generated and label interferograms, $N_x$ and $N_y$ are the pixels length in the $x$ and $y$ axis of input image. $\Delta\theta$ is the phase shift error. The loss function of the generator network contains cross entropy, RMSE and phase shift error to minimize the RMSE between the generated and label interferograms and the error of phase shift.

**Simulation results**

To verify the feasibility and precision of our method, simulation initial interferograms are used to perform deep phase shift and phase retrieval. Simulation initial interferogram is out of the training data. The DPS-net is trained to generate multiple interferograms with arbitrary phase shifts, such as 2π/3, -3π/2, π/2, π and 3π/2. The corresponding generated interferograms are shown as in Fig. S2-S3 (in Supplement 1). To analyze the accuracy of phase shift, the generated phase shift interferograms is compared with the simulation phase shift interferograms, as shown in Table S1-S2 (in Supplement 1). The phase shifts are calculated with AIA algorithm [33]. The standard deviation of phase shifts is 0.0091 rad. The average RMSEs of generated phase shifting interferograms is 1.922 gray level. They show a high accuracy. The initial interferogram and generated interferograms with π/2, π and 3π/2 phase shift are used to perform phase retrieval by using four-step algorithm, as $\varphi = \arg(I_0 - I_\pi, I_{3\pi/2} - I_{\pi/2})$. The results of phase shift and phase retrieval and the corresponding error distribution are shown as in Fig. 3. The reference phase and corresponding phase retrieval results are used to calculate the RMSEs of phase retrieval results. The average RMSEs of phase retrieval is 0.0962 rad. It still shows a high accuracy.

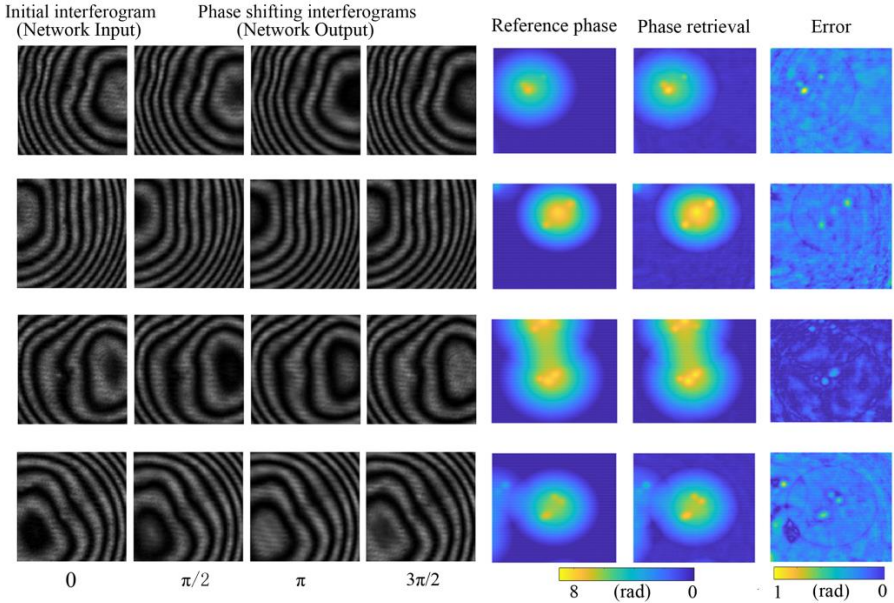

Fig. 3 The simulation results of deep phase shift and phase retrieval

**Experimental results**

To verify the feasibility and precision of our method, experimental recorded interferogram are used to perform deep phase shift and phase retrieval. Hemispherical polyvinyl chloride pellets and Hela cells are measured by using Mach-Zehnder interference system. The reference phase is calculated from 30 frames phase shifting interferogram by using piezoelectric transducer (PZT) phase shifter. In the experimental verify section, only one interferogram is input the trained DPS-net. And three interferograms with $\pi/2$, $\pi$ and $3\pi/2$ phase shift can be obtained. And the phase shifts are calculated with AIA algorithm, phase retrieval results are calculated with four-step algorithm. The phase shifting interferograms and phase retrieval results of hemispherical polyvinyl chloride pellets are shown as in Fig. 4. The standard deviation of phase shifts is 0.0185 rad. And the RMSE of phase retrieval result is 0.1183 rad.

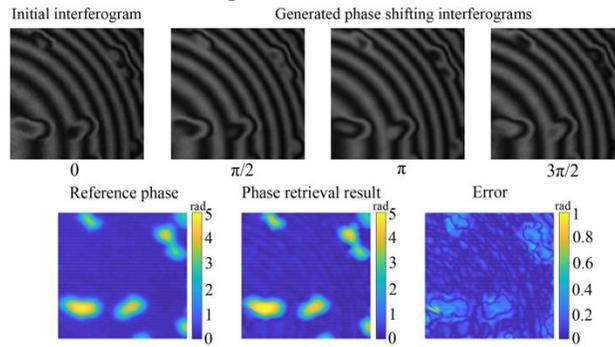

Fig. 4 The generated phase shifting interferograms and phase retrieval results of hemispherical polyvinyl chloride pellets

The other experiment samples are the Hela cells. The phase shifting interferograms and phase retrieval results of Hela cells are shown as in Fig. 5. And The standard deviation of phase shifts is 0.0263 rad. The average RMSEs of phase retrieval is results is 0.1608 rad. Compare with the phase retrieval results of hemispherical polyvinyl chloride pellets and simulation results. The RMSEs of Hela cells phase retrieval have a slight increase, but we can still obtain the Hela cells quantitative phase image from single interferogram in in-line holography geometry with high accuracy.

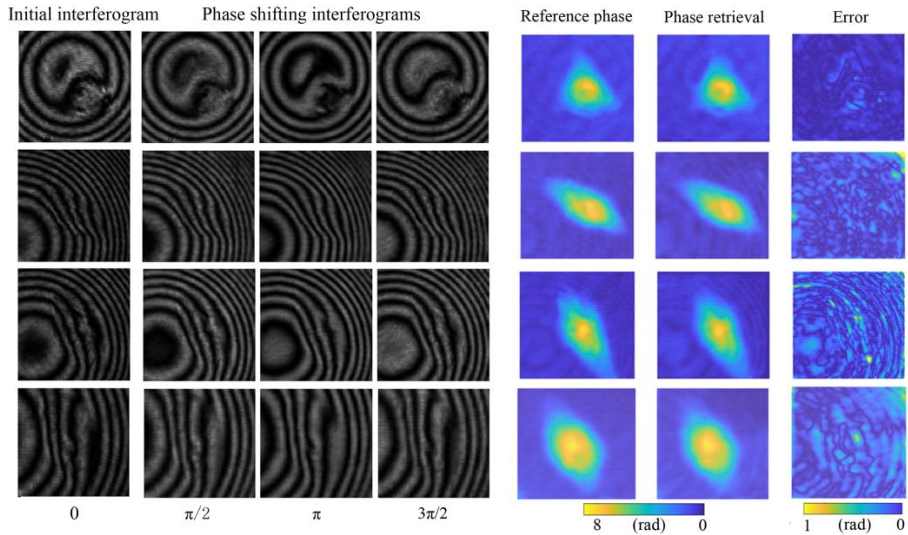

Fig. 5 The generated phase shifting interferograms and phase retrieval results of Hela cells.

## 4. Conclusion

In this paper, we demonstrate that a DPS-net can be trained to perform the phase shift in in-line holography. The trained DPS-net can be used to predict multiple interferograms with arbitrary phase shift from a single interferogram as a full digital phase shifter. The quantitative phase image of hemispherical polyvinyl chloride pellets and Hela cells sample are obtained from only a single interferogram with high accuracy. The simulation and experimental results demonstrate that the feasibility and high accuracy of our method. It is worth mentioning that deep learning can be used to replace the entitative phase shift operation, and improve the full-field quantitative phase imaging technology.